\documentclass[aps,prl,floatfix,twocolumn]{revtex4}
\usepackage{dcolumn}
\usepackage{bm}
\usepackage[dvips]{epsfig}
\usepackage{graphicx} \usepackage{amsmath} \usepackage{amssymb}
\newcommand{\comment}[1]{}
\newcommand{\BEQ}{\begin{equation}}
\newcommand{\EEQ}{\end{equation}}
\newcommand{\BEA}{\begin{eqnarray}}
\newcommand{\EEA}{\end{eqnarray}}
\renewcommand{\d}{{\rm d}}

\renewcommand{\L}{{\rm L}}
\newcommand{\R}{{\rm R}}
\newcommand{\W}{{\rm W}}

\renewcommand{\k}{{\rm k}}
\newcommand{\F}{{\bf F}}

\newcommand{\x}{\hat{x}}
\newcommand{\E}{\hat{E}}
\newcommand{\y}{\hat{y}}

\newcommand{\lar}{\leftarrow}

\begin{document}

\title{Adaptive heat engine}

\author{A.E. Allahverdyan$^1$, S.G. Babajanyan$^1$, 
N.H. Martirosyan$^1$, A.V. Melkikh$^2$}
\affiliation{$^1$Yerevan Physics
  Institute, Alikhanian Brothers Street 2, Yerevan 375036, Armenia}
\affiliation{$^2$Ural Federal University,
  Mira Street 19, Yekaterinburg 620002, Russia}





\begin{abstract}
  A major limitations for many heat engines is that their functioning
  demands on-line control, and/or an external fitting between
  environmental parameters (e.g. temperatures of thermal baths) and
  internal parameters of the engine. We study a model for an adaptive
  heat engine, where|due to feedback from the functional part|the
  engine's structure adapts to given thermal baths. Hence no on-line
  control and no external fitting are needed. The engine can employ
  unknown resources; it can also adapt to results of its own
  functioning that makes the bath temperatures closer. We determine
  thermodynamic costs of adaptation and relate them to the prior
  information available about the environment. We also discuss
  informational constraints on the structure-function interaction that
  are necessary for adaptation.

\end{abstract}






\maketitle

{\it Introduction.} Heat-engines drove the Industrial Revolution and
their foundation, {\it viz.} thermodynamics, became one of the most
successful physical theories \cite{callen}. Its extensions to
stochastic \cite{sekimoto,seifert} and quantum domain \cite{mahler}
led to new generations of heat engines
\cite{seifert,mahler,debashish,scovil,prokhor,
  kosloff,tannor,uzdin,johal,arm,pop,barba,scully}. As everyone could
observe, the work-extraction function of macroscopic heat-engines
requires external on-line control, e.g. the specific sequence of
adiabatic and isothermal processes for the Carnot cycle
\cite{callen,kauffman}. Smaller engines may not demand on-line
control, i.e. they are autonomous \cite{pop,barba}, but they do demand
fitting between internal and environmental parameters
\cite{scovil,prokhor,kosloff,tannor,uzdin,johal,arm}, e.g. because for
fixed environment (thermal baths) there are internal parameters, under
which the machine acts as a heat-pump or refrigerator performing tasks
just opposite to that of heat-engine. Such fitted engines are
susceptible to environmental changes, e.g. when the bath temperatures
get closer due to the very engine functioning. Car engines treat this
problem by abandoning the partially depleted fuel (i.e. the hot bath),
and using fresh fuel.

Here we study a rudimentary model of autonomous, adaptive heat
engine. Adaptive means that the engine can work for a sufficiently
general class of environments, i.e. it does need
neither on-line control, nor an externally imposed fitting between its
internal parameters and the bath temperatures. In particular, the
engine can adapt to the results of its own functioning.

The major biophysical heat engine, {\it viz} photosynthesis|which
operates between the hot Sun temperature and the low-temperature Earth
environment \cite{mex}|does have adaptive features that allow its
functioning under decreased hot temperature (partial shadowing) or
increased cold temperature (hot whether)
\cite{igor,adaptive_photo_s}. Hence adaptive engines can be useful for
fueling devices employing unknown or scarce resources. They are
already employed in engineering for photovoltaic engines that can
operate under partial shadow \cite{adaptive_photo_v}.

Recently, several physical models concentrated on adaptive sensors
(learning and memorizing environmental changes), adaptive transport
models {\it etc}
\cite{igor,tamar,kaneko,sartori,adaptive,barato,bo,sartori_prl}. These
studies clarified thermodynamic costs of adaptation scenarios
\cite{igor,sartori,adaptive,barato,bo,sartori_prl}. Other research
lines related adaptation with (poly)homeostasis \cite{gros} and models
of artificial life \cite{kauffman,virgo}.

For analyzing the adaptation and its costs for heat engines, we need a
tractable and realistic model that is much simpler than its prototypes
in photovoltaics or photosynthesis. The model ought to consist of
functional and structural parts. For the former we choose one of the
most known models of quantum/stochastic thermodynamics
\cite{scovil,prokhor,kosloff,tannor,uzdin}.

\comment{
Motivations for studying autonomous, adaptive engines (artificial
life, enzyme functioning, adaptive sensors). Homeostasis (stability
with respect to environment), robustness (passive stability), active
seeking for energy sources, employing feedback from its own
structures.

One motivation for studying autonomous, universal engines comes from
nano-technologies. Nano-robots are to be fueled autonomously,
i.e. from small heat-engines. Making such engines universal is
certainly useful for polyfunctional robots.

Another motivation comes from the metabolism-first scenario of the
emergence of life. Within this scenario the advanced forms of life
from simple heat or chemical engines. Such engines are expected to be
autonomous (if we believe that the life emerged without external
design).

Thus autonomous and universal are closely related: the first forbids
on-line control of engine-functioning, while the second refers to
initial (and constant) values of internal parameters.

}

{\it The functional degree of freedom} of the model engine has three
states: $i=1,2,3$ \cite{scovil,prokhor,kosloff,tannor,uzdin}. Its
dynamics is described by a Markov master equation \cite{kampen}: \BEA
\label{1}
\dot{p}_i\equiv {\d p_i}/{\d t}= {\sum}_{j}[\rho_{i\leftarrow
  j}p_j-\rho_{j\leftarrow i}p_i], \quad i,j=1,2,3, \EEA where $p_i$ is
the probability of $i$, and $\rho_{i\leftarrow j}$ is the transition
rate from $j$ to $i$. The stationary solution of (\ref{1}) is
\begin{eqnarray}
  \label{77}
  p_1=\frac{1}{{\cal Z}}
[\rho_{1\lar 2}\rho_{1\lar 3}+\rho_{3\lar
  2}\rho_{1\lar 3}+\rho_{2\lar 3}\rho_{1\lar 2}],
  \\
  \label{777}
  p_2=\frac{1}{{\cal Z}}[\rho_{2\lar 1}\rho_{2\lar 3}+\rho_{3\lar
    1}\rho_{2\lar 3}+\rho_{2\lar 1}\rho_{1\lar 3}],
  \\ 
  p_3=\frac{1}{{\cal Z}}[\rho_{3\lar 1}\rho_{3\lar 2}+
  \rho_{2\lar 1}\rho_{3\lar 2}+\rho_{1\lar 2}\rho_{3\lar 1}], 
  \label{7777}
\end{eqnarray}
where ${\cal Z}$ ensures $p_1+p_2+p_3=1$. We assume that each
transition $i\leftrightarrow j$ couples with the bath at inverse
temperatures $\beta_{ij}=\beta_{ji}$. The detailed balance reads
\cite{kampen}:
\begin{gather}
  \label{eq:2}
  \rho_{i\lar j}\, e^{-\beta_{ij}E_j}=  \rho_{j\lar i}\, e^{-\beta_{ij}E_i}, \qquad
\beta_{ij}=\beta_{ji},
\end{gather}
where $E_i$ is the energy of $i$. One of the baths has infinite
temperature: $\beta_{12}=0$. It is standardly associated with a
work-source, because due to $\d S_{12}=\beta_{12}\d Q_{12}=0$ it
exchanges energy $\d Q_{12}\not=0$ at zero entropy increase $\d
S_{12}=0$. Effectively large temperatures arise naturally in
biomolecular systems due to the stored energy, which|when recalculated
in terms of temperature|is some 20--50 times larger than the room
temperature \cite{stored}.

The model (\ref{1}, \ref{eq:2}) was introduced and studied
in the quantum setting \cite{scovil,prokhor,kosloff,tannor,uzdin}, as
a model for maser. Closely related models were studied recently in
the context of photovoltaic engines \cite{scully}.

The average energy conservation in the stationary regime reads
from (\ref{1}): $\sum_{k=1}^3\dot{p}_kE_k =J_{31}+J_{32}+J_{21}=0$, where
$J_{i>j}$ is the energy current from the bath that drives the
transition $i\leftrightarrow j$:
\begin{gather}
  \label{eq:1}
J_{i>j}=(E_i-E_j)(\rho_{i\lar j}p_j-\rho_{j\lar i}p_i).
\end{gather}
$J_{i>j}$ is positive when the energy comes out from the bath. Using
(\ref{77}--\ref{7777}) we get in the stationary state
\begin{gather}
  \label{eq:5}
J_{21}=\frac{\E_2}{{\cal Z}}\, \rho_{1\lar 3}\,\rho_{3\lar 2}\,\rho_{1\lar
  2}\left[
1-e^{  (\beta_{32}- \beta_{31})\E_3-\beta_{32}\E_2  }
\right],\\
  \label{eq:55}
J_{31}=-{\E_3J_{21}}/{\E_2},~~
J_{32}={(\E_3-\E_2)J_{21}}/{\E_2},\\
\E_2\equiv E_2-E_1, \qquad \E_3\equiv E_3-E_1.
\label{bela}
\end{gather}
The heat-engine functioning is defined as [cf. (\ref{eq:2}, \ref{eq:1})]
\begin{eqnarray}
  \label{eq:4}
0>  J_{21}=(E_2-E_1)(p_1-p_2)\rho_{1\lar 2},
\end{eqnarray}
i.e. the energy goes to the work-source. 
Note that (\ref{eq:4}) relates to population inversion between energy
levels $E_1$ and $E_2$. Using (\ref{eq:5}) we write (\ref{eq:4}) as
\begin{gather}
  \label{eq:7}
\E_2[ (1-\vartheta)\E_3-\E_2]>0, \quad 
\vartheta\equiv\beta_{31}/\beta_{32}.  
\end{gather}
Eq.~(\ref{eq:7}) demands different temperatures: $\beta_{32}\not=
\beta_{31}$. It also demands tuning between the energies $\E_2$,
$\E_3$ and $\vartheta$: it is impossible to hold (\ref{eq:7}) for a
wide range of $\vartheta$ by means of constant $\E_2$ and $\E_3$;
e.g. if (\ref{eq:7}) holds for $1>\vartheta$ due to $\E_3>\E_2>0$,
then it is violated for $1-\vartheta <\frac{\E_2}{\E_3}$. Tuning
is necessary, since for suitable values of $\E_2$ and $\E_3$, the
machine can function also as a refrigerator (i.e. $J_{21}>0$ and
$J_{32}>0$ for $\beta_{32}>\beta_{31}$) or as a heat-pump.

\comment{Not all non-equilibrium features are fragile in this sense:
  consider e.g. the cyclic transformation of the probability between
  states $1\lar 2\lar 3\lar 1$. This is determined by positivity of
  three probability currents $I_{1 \leftarrow 2}$, $I_{3\leftarrow 1}$
  and $I_{2\leftarrow 3}$, where $I_{i\leftarrow j}=\rho_{i\lar
    j}p_j-\rho_{j\lar i}p_i$ is read-off from
  (\ref{eq:1}--\ref{eq:55}). The cycle holds for
  $(1-\vartheta)\E_3>\E_2$, which is valid for all $\vartheta$ when
  taking $\E_3=0$ and $0>\E_2$.}


{\it The structural degree of freedom} $x$ is continuous, since it
should ensure adaptation to continuous environmental
variations. $x$ governs the behavior of interaction energies
$E_i(x)$ between $x$ and $i$. The joint probability $p_i(x,t)$ of $x$
and $i$, $\int\d x\sum_i p_i(x,t)=1$, evolves via the Fokker-Planck
plus master equations [cf. (\ref{1})] \cite{kampen}:
\begin{eqnarray}
  \label{eq:8}
  \dot{p}_i(x,t)&=&{\sum}_{j}[\rho_{i\lar j}(x)p_j(x,t)-\rho_{j\lar
    i}(x)p_i(x,t)] ~~~~\\
&+&\frac{1}{\gamma}
  \partial_x[p_i(x,t)E'_i(x)]+D \partial^2_x p_i(x,t),
  \label{eq:9}
\end{eqnarray}
where $ i,j=1,2,3$, $E'_i(x)\equiv\frac{\d E_i(x)}{\d x}$, $\gamma>0$
is the damping constant, and $D>0$ is the diffusion constant.
$\rho_{i\lar j}(x)$ is specified in (\ref{eq:22}); it holds
(\ref{eq:2}) with $E_i\to E_i(x)$ and $E_j\to E_j(x)$. 

Eqs.~(\ref{eq:8}, \ref{eq:9}) is well-known in chemical and biological
applications \cite{zwanzig,hans,leh,agmon_h,blum,conformon,christo}.
The limit when $E_i(x)=E(x)$ does not depend on $i$ refers to a
dynamic disorder \cite{zwanzig,hans,leh,agmon_h}. It was applied to a
variety of problems including the conformational dynamics of enzymes
and ion channels \cite{zwanzig,hans,leh,agmon_h}. These systems also
provide examples, where the dependence of $E_i(x)$ on $i$ (feedback)
is essential \cite{leh,blum,conformon,christo}.

We assume in (\ref{eq:8}) that $x$ is slow: $\frac{1}{\gamma},\,D\ll
\rho_{i\lar j}(x)$. This limit is implemented by introducing in
(\ref{eq:9}, \ref{eq:8}) the conditional probability $p_{i|x}(t)$
\cite{a+n},
\begin{eqnarray}
p_i(x,t)=p_{i|x}(t)p(x,t),\,
\int\d x \, p(x,t)=1, \, \sum_i p_{i|x}(t)=1,\nonumber
\end{eqnarray}
and collecting fast terms:
\begin{eqnarray}
  \label{eq:11}
 \dot{p}_{i|x}={\sum}_{j}[\rho_{i\lar j}(x)p_{j|x}-\rho_{j\lar i}(x)p_{i|x}].
\end{eqnarray}
Slow terms are found from (\ref{eq:9}, \ref{eq:11}) by summing
over $i$:
\begin{eqnarray}
  \dot{p}(x,t)=
\frac{1}{\gamma}
  \partial_x[p(x,t)\sum_i p_{i|x} E'_i(x)]+D\partial^2_x p(x,t). 
  \label{eq:111}
\end{eqnarray}
Since $i$ is fast, $p_{i|x}$ in (\ref{eq:111}) can be taken as
time-independent, i.e. $p_{i|x}$ is found from (\ref{77}--\ref{7777})
upon replacing there $\rho_{ij}\to \rho_{ij}(x)$ \cite{a+n}. The
stationary probability of $x$ is found from (\ref{eq:111}) via the
zero-current condition $\frac{p(x)}{\gamma}\sum_i p_{i|x}
E'_i(x)+D\partial_x p(x)=0$:
\begin{eqnarray}
  \label{eq:12}
  p(x)\propto e^{-\Psi(x)/(\gamma D)}, \quad \Psi'(x)\equiv{\sum}_{i=1}^3
  p_{i|x} E'_i(x),
\end{eqnarray}
where $\Psi(x)$ is an effective potential of $x$, and
$\Psi'(x)=\frac{\d\Psi}{\d x}$.

{\it Adaptation.}  Naturally, the energies $E_i(x)$ do not depend on
$\beta_{31}$ and $\beta_{32}$. We choose $E_i(x)$ such that two
conditions hold. First, $\Psi(x)$ has a unique minimum $\x$:
\begin{eqnarray}
  \label{eq:13}
&&  \Psi'(\x)={\sum}_{i=1}^3 p_{i|\x}
E'_i(\x)=0, \quad 
\Psi''(\x)>0.
 \end{eqnarray}
$\x$ is the unique maximally probable value of $x$; cf. (\ref{eq:12}). 

\begin{figure}
\includegraphics[width=7cm]{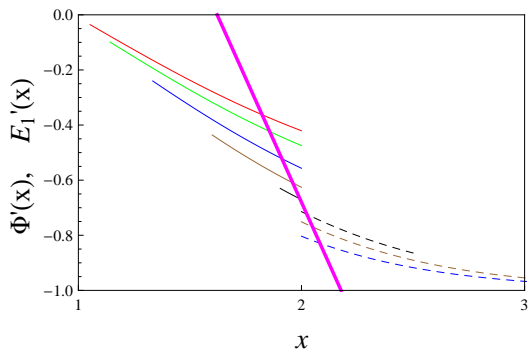} 
\caption{$\Phi'(x)$ given by (\ref{eq:28}, \ref{eq:22}) with
  $f[y,\beta]=e^{y/2}$ for varying $\beta_{31}$ and fixed
  $\beta_{32}=1$ (first type of varying environment).  We assume
  $\E_3(x)=-x$, $\E_2(x)=x-2$, and (\ref{kaa}) holds for $x>2$ if
  $\vartheta>2$, for $\frac{2}{2-\vartheta}>x>2$ if $1<\vartheta<2$,
  and for $\frac{2}{2-\vartheta}<x<2$ if $\vartheta<1$. $\Phi'(x)$ is
  shown for various $\vartheta=\beta_{31}/\beta_{32}$ and those $x$
  that support (\ref{kaa}): $\vartheta=0.1$ (red curve),
  $\vartheta=0.25$ (green), $\vartheta=0.5$ (blue), $\vartheta=0.75$
  (brown), $\vartheta=0.95$ (black), $\vartheta=1.2$ (black-dashed),
  $\vartheta=1.5$ (brown-dashed), $\vartheta=2.5$ (blue-dashed). The
  magenta curve shows $-E_1'(x)$, where $E_1'(x)=1.8(x
  -2)+0.680289$. Intersections of $-E_1'(x)$ with $\Phi'(x)$ determine
  $\x$.  Eqs.~(\ref{eq:29}) hold for all $\vartheta$. }
\label{fig1}
\end{figure}

\begin{figure}
\includegraphics[width=7cm]{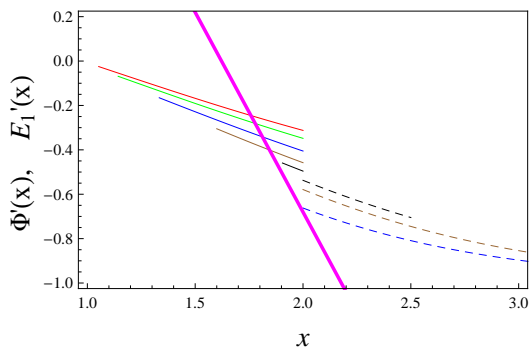}
\caption{The same as in Fig.~\ref{fig1}, but for $\beta_{32}=0.7$. It
  is seen that for $\vartheta$ close to $1$, no heat-engine
  functioning exists ((\ref{kaa}) does not hold), i.e. the magenta
  curve does not cross the curves with $\vartheta=0.95$,
  $\vartheta=1.2$ and $\vartheta=1.5$.  }
\label{fig1_1}
\end{figure}

Second, the heat-engine condition $J_{21}(x)<0$ holds in a
vicinity of the maximally probable value $\x$ [cf. (\ref{eq:7})]:
\begin{eqnarray}
  \label{kaa}
\E_2(x)[ (1-\vartheta)\E_3(x)-\E_2(x)]>0, ~~ x\simeq \x,
\end{eqnarray}
where $\E_i(x)=E_i(x)-E_1(x)$; cf. (\ref{bela}). Eqs.~(\ref{eq:13},
\ref{kaa}) imply feedback adaptation: if $\beta_{31}$ or $\beta_{32}$
change, the system is not anymore in a stationary state. It then goes
to new stationary state, where the heat-engine function most probably
holds.

\comment{
Taking $D\gamma$ sufficiently small ensures from (\ref{eq:12},
\ref{kaa}) the work extraction in average: $\int\d x\,
p(x)\,J_{21}(x)<0$.
}

Note that $x$ is continuous, because we shall assume that $\beta_{31}$
and $\beta_{23}$ change in continuous domains. Using here a
feed-forward (instead of feedback) control does not lead to
adaptation, because it influences {\it only} the diffusion constant
$D$; see sect. 1 of suppl. material.


To get a general method for studying (\ref{eq:13}, \ref{kaa}), we
focus on the following class of transition rates $\rho_{ij}$
[cf. (\ref{1}, \ref{eq:2})]:
\begin{eqnarray}
  \label{eq:22}
  \rho_{ij}(x)=f[\,\beta_{ij} (E_j(x)-E_i(x)),\beta_{ij}\,], ~~~ 
\beta_{ij}=\beta_{ji},
\end{eqnarray}
where $f[y,\beta]$ holds (\ref{eq:2}).
Eq.~(\ref{eq:22}) implies that $\rho_{ij}(x)$ and the stationary
probabilities $p_{i|x}$ depend only on $\E_3(x)$ and $\E_2(x)$;
cf. (\ref{eq:22}, \ref{bela}). This holds in the high-temperature
limit for any $\rho_{ij}$; see (\ref{eq:2}). Two other examples of
(\ref{eq:22}) is the Kramers' rate
$f[y,\beta]=e^{\beta\delta+{\rm min}(y,0)}$, where
$\delta$ is the barrier height \cite{kampen}, and
$f[y,\beta]=e^{y/2}$ that corresponds to the discrete-space
Fokker-Planck equation \cite{agmon_h}.  \comment{We get from
  (\ref{eq:2}) within the order ${\cal O}(\beta_{32})$ and ${\cal
    O}(\beta_{31})$
\begin{eqnarray}
  \label{eq:07}
 p_{1|x}=\frac{1}{9}[3+\beta_{31}\E_3(x)],\\
  \label{eq:17}
 p_{2|x}=\frac{1}{9}[3+\beta_{32}(\E_3(x)-\E_2(x))],\\
  \label{eq:1717}
 p_{3|x}=\frac{1}{9}[3+\beta_{32}(\E_2(x)-\E_3(x))-\beta_{31}\E_3(x)].
\end{eqnarray}}
We use $\sum_{i=1}^3p_{i|x}=1$ and define from (\ref{eq:13}, \ref{eq:22}) 
\begin{eqnarray}
  \label{eq:28}
\Phi'(x)  = {\sum}_{i=2}^3 p_{i|x} \E'_i(x),
\end{eqnarray}
where the analogues of (\ref{eq:13}) read
\begin{eqnarray}
  \label{eq:29}
\Phi'(\x)=-E'_1(\x), \qquad
\Phi''(\x)>-E''_1(\x).
\end{eqnarray}
For given $\E_3(x)$ and $\E_2(x)$, $\Phi'(x)$ does not depend on
$E_1(x)$; see (\ref{77}--\ref{7777}). Hence one can study $\Phi'(x)$
for given $\E_3(x)$ and $\E_2(x)$ [cf. (\ref{kaa})] and for different
values of $\beta_{31}$ and $\beta_{32}$. Then one can define $\x$ via
(\ref{eq:29}) by choosing a suitable $E_1(x)$ that does not depend on
$\beta_{31}$ and on $\beta_{32}$.

Since (\ref{kaa}) should hold for $\vartheta\to 1$, there exists
$x_0$ such that $\x\to x_0$ for $\vartheta\to 1$, and
$\E_2(x_0)=0$. In the vicinity of $x_0$, $\E_3(x)$ is either finite or
goes to zero slower than $\E_2(x)$, so that (\ref{kaa}) still holds
for $\vartheta\to 1$ and $\x\to x_0$.

Let us first assume that one temperature (say $\beta_{31}$) takes
arbitrary positive values, while another one ($\beta_{32}$) is
fixed. Adaptation is necessary here, since
$\vartheta=\beta_{31}/\beta_{32}$ is an arbtrary positive number,
hence (\ref{kaa}) cannot be valid for $x$-independent $E_i$. Now
(\ref{eq:13}, \ref{kaa}) for adaptation can be satisfied; see
Fig.~\ref{fig1} for the simplest but representative choice, where
$E_i(x)$ are parabolic functions of $x$. This choice of $E_i(x)$ is
realistic \cite{leh,blum,conformon,christo}.


Since the validity domain (\ref{kaa}) of the heat-engine shrinks to a
point for $\vartheta\to 1$, we need progressively smaller values of
$D\gamma$ in (\ref{eq:12}) for ensuring the average work-extraction
\begin{eqnarray}
  \label{eq:40}
  \langle J_{21}\rangle \equiv \int\d x\, p(x)\,J_{21}(x)<0
\end{eqnarray}
for $\vartheta\to 1$. If the diffusion of $x$ is caused by an
equilibrium bath, we get $D\gamma=T$ \cite{kampen}, and the
temperature $T$ of this bath should be sufficiently low for
(\ref{eq:40}) to hold. If this is the lowest temperature, there is a
heat current towards it tending to increase it. Hence this low
temperature is a thermodynamic resource. For a given $D\gamma$, there
is a vicinity of $\vartheta= 1$, where no work is extracted in
average: $\langle J_{21}\rangle>0$.

\begin{figure}
\includegraphics[width=7cm]{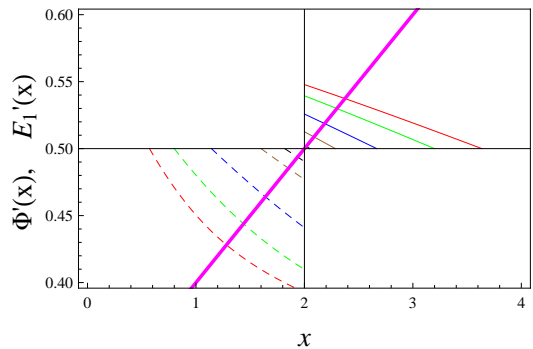}
\caption{Adaptation for a negative friction $\gamma<0$, $\beta_{32}=1$
  and varying $\beta_{13}$. The same parameters as in Fig.~\ref{fig1},
  but now $\E_3(x)=x/2$. Conditions (\ref{kaa}) amount to
  $\frac{4}{1+\vartheta}>x>2$ if $\vartheta <1$, and to
  $\frac{4}{1+\vartheta}<x<2$ if $\vartheta >1$. $\Phi'(x)$ is shown
  for: $\vartheta=0.1$ (red curve), $\vartheta=0.25$ (green),
  $\vartheta=0.5$ (blue), $\vartheta=0.75$ (brown), $\vartheta=0.95$
  (black), $\vartheta=1.2$ (black-dashed), $\vartheta=1.5$
  (brown-dashed), $\vartheta=2.5$ (blue-dashed), $\vartheta=4$
  (green-dashed), $\vartheta=6$ (red-dashed). The magenta curve shows
  $-E_1'(x)$, where $E_1'(x)=-0.1(x -2)-0.5$. Eq.~(\ref{eq:2929})
  holds.  }
\label{fig2}
\end{figure}

\begin{figure}
\includegraphics[width=7cm]{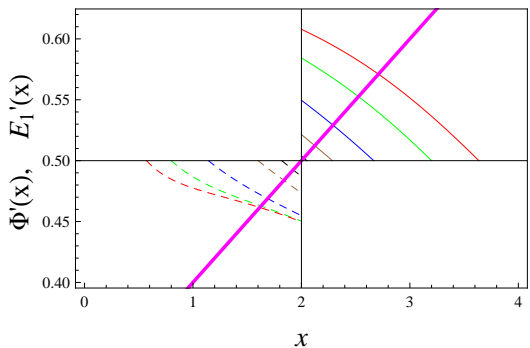}
\caption{The same as in Fig.~\ref{fig2}, but for
  $\beta_{32}=3$. Eqs.~(\ref{eq:2929}, \ref{kaa}) hold.  }
\label{fig2_1}
\end{figure}

Consider now a general situation, where both $\beta_{31}$ and
$\beta_{32}$ vary. Fig.~\ref{fig1_1} shows that the set-up which
worked for a fixed $\beta_{32}$ does not apply here. Now condition
$\Phi'(\x)=-E'_1(\x)$ in (\ref{eq:29}) implies such a behavior for
$\Phi'(\x)$ under $\beta_{31}\approx \beta_{32}$ that the second
condition $\Phi''(\x)>-E''_1(\x)$ in (\ref{eq:29}) cannot hold;
e.g. because $\Phi'(\x)$ has the shape shown in Figs.~\ref{fig2} and
\ref{fig2_1}.  This fact is shown in sect. 2 of 
suppl. material. The only possibility to recover the adaptive heat
engine function is to assume that $x$ is subject to an external force
that adds to RHS of (\ref{eq:111}) a contribution making the effective
friction negative: $\gamma<0$. Then $D\gamma<0$ in (\ref{eq:12}), and
the most probable $\x$ means that inequalities in (\ref{eq:13}) and
(\ref{eq:29}) are reversed.  Now the adaptation conditions are
(\ref{kaa}) and [instead of (\ref{eq:29})]
\begin{eqnarray}
  \label{eq:2929}
\Phi'(\x)=-E'_1(\x), \qquad
\Phi''(\x)<-E''_1(\x).
\end{eqnarray}
These conditions can be satisfied, as illustrated in Figs.~\ref{fig2}
and \ref{fig2_1}. Now $x$ should change in a bounded domain; otherwise
for the natural shape of energies ($E_i(x)\to \infty$ for
$x\to\pm\infty$) one gets a non-normalizable $p(x)$ in
(\ref{eq:12}). Note that for $\gamma<0$ and $D>0$, (\ref{eq:111}) does
predict relaxation to (\ref{eq:12}), i.e. the negative-friction
situation is stable.  The external force that simulates the negative
friction does dissipate energy with a constant rate
$\Pi=\frac{1}{|\gamma|}\int \d x\, p(x)[\Psi'(x)]^2$; see sect. 3 of
suppl. material. We still demand a sufficiently small $|\gamma|D$
to ensure $\langle J_{21}\rangle <0$. For a small $|\gamma|D$ we get
$\Pi\simeq |\Psi''(\x)|D$. This energy dissipation is another cost for
adaptation; see \cite{sartori,adaptive,barato,bo,sartori_prl} for
related results. A negative friction is known in several classes of
active (non-equilibrium) systems \cite{samo,wiki,starr,brown}.  Two
examples relevant to our situation is the negative resistance of
electric circuits \cite{wiki} and negative viscosity of driven fluids
\cite{starr}.


Above examples of adaptation are obtained for $E_i$ being correlated
random variables, $P(E_1,E_2,E_3)=\int\d x\, p(x)\prod_{k=1}^3
\delta(E_k-E_k(x))$, and for specific forms of $E_i(x)$. To see why
such assumptions are necessary, take an extreme case, where the
probability density of energies $\Pi(E_1,E_2,E_3)$ is {\it
  non-informative} in terms of the Bayesian statistics
\cite{jaynes,johal}. Since we do not have prior expectations about
correlations between random variables $E_1$, $E_2$ and $E_3$, they are
taken as independent $\Pi(E_1,E_2,E_3)=\prod_{k=1}^3 \Pi(E_k)$
\cite{jaynes}. Also, since $E_1$ can assume either sign, the
non-informative density $\Pi(E_k)$ is the homogeneous one:
$\Pi(E_k)={\rm const}$ \cite{jaynes}. Employing this density we
calculate from (\ref{eq:7}) that the probability of the heat-engine
functioning is low $\frac{1-{\rm
    min}(\vartheta,\frac{1}{\vartheta})}{3}<\frac{1}{3}$.  Hence the
current $J_{21}$ averaged over $\Pi(E_1,E_2,E_3)$ is positive, i.e.
the machine does not function as a heat-engine; cf. (\ref{eq:5}).
Thus the coupling between structure and function, which is encoded in
$P(E_1,E_2,E_3)$ must be informative.

In sum, we studied a model for an adaptive heat engine that can
function under scarce or unknown resources. Several physical
limitations for the adaptation concept were uncovered; they relate to
the prior information available about the environment.  One problem
generated by this research is that in the model the resources needed
for adaptation are detached from the work extracted by the
engine. Employing the extracted work for ensuring the adaptation will
make the situation more interesting.

This work is partially supported by COST MP1209 and by the ICTP
through the OEA-AC-100.

\comment{
\begin{gather}
  \label{eq:16}
  P(E_1,E_2,E_3)=\pi (E_1)\pi (E_2)\pi (E_3), \\
 \pi(x)=\frac{1}{2A}\theta[x+A]\theta[A-x],  ~~ \int\d x\, \pi(x)=1,~~
  \label{eq:14}
\end{gather}
where $\theta[x]$ is the step function: $\theta[x\leq 0]=0$,
$\theta[x> 0]=1$, and where $A>0$ serves for regularizing the
homogeneous density, it will drop out from the final result.}



\section{Supplementary material}

\subsection{1. Feed-forward control}

Feed-forward amounts to no direct interaction between the functional
degree of freedom $i$ and the structural degree of freedom $x$. But
now $x$ couples directly to the baths at temperatures
$T_{31}=1/\beta_{31}$ and $T_{32}=1/\beta_{32}$, i.e. we try to
implement temperature sensors via $x$.
This means taking in (\ref{eq:9}) of the main
text $E_i(x)=E(x)$ and adding there the following term
\begin{eqnarray}
  \label{eq:37}
\sum_{\alpha={31,32}} \gamma^{-1}_{\alpha}[
\partial_x[p_i(x,t)E'(x)] + T_\alpha\partial^2 p_i(x,t)].   
\end{eqnarray}
Instead of (\ref{eq:12}) [of the main text] we get for the stationary
probability
\begin{eqnarray}
  \label{eq:38}
\bar{p}(x)\propto e^{-E(x)/\widetilde{D}}, \qquad \widetilde{D}=
\frac{D+\sum_{\alpha={31,32}} \gamma^{-1}_{\alpha}
  T_{\alpha}}{\gamma^{-1}+\sum_{\alpha={31,32}} \gamma^{-1}_{\alpha}
},   
\end{eqnarray}
where $E(x)$ does not depend on $\beta_{31}$ and
$\beta_{32}$. Hence no adaptation is possible.

\subsection{2. Derivation of the no-adaptation condition}

If both $\beta_{31}$ and $\beta_{32}$ can vary, 
the adaptation condition $\Phi'(\x)=-E'_1(\x)$
[see (\ref{eq:29}) of the main text] should hold for all
$\beta_{31}$ and $\beta_{32}$. In particular, this means that|since
$E_1'(x)$ does not depend on $\beta_{31}$ and $\beta_{32}$|
  $\Phi'(x_0)\left|_{\beta_{31}=\beta_{32}=\beta}\right.$ should not
  depend on $\beta_{31}=\beta_{32}=\beta$, where $\E_2(\x=x_0)=0$.  
Now (\ref{77}--\ref{7777}) of the main text
  show that $p_{i|x_0}$ are at equilibrium for
  $\beta_{31}=\beta_{32}=\beta$ and 
  \begin{gather}
    \label{eq:25}
p_{1|x_0}=p_{2|x_0}=\frac{1}{Z}, \qquad 
p_{3|x_0}=\frac{e^{-\beta \E_3}}{Z}.
\end{gather}
Hence we get
  \begin{gather}
  \label{eq:32}
\Phi'(x_0)\left|_{\beta_{32}=\beta_{31}=\beta}\right.=
\frac{\E_2'(x_0)+\E_3'(x_0)e^{-\beta \E_3(x_0)}}{2+e^{-\beta \E_3(x_0)}}.
\end{gather}
This expression is not a function of $\beta$ only for
\begin{eqnarray}
  \label{eq:21}
  \E_2'(x_0)=2\E_3'(x_0), 
\end{eqnarray}
where
\begin{eqnarray}
  \label{eq:42}
  \Phi'(x_0)\left|_{\beta_{31}=\beta_{32}=\beta}\right.=
\E_3'(x_0).
\end{eqnarray}
Using (\ref{eq:21}) we find $\Phi'(x)$ for
$x\approx x_0$:
\begin{gather}
  \Phi'(x)=   \Phi''(x_0)
(x-x_0)+\E_2'(x_0)p_{2|x_0}+\E_3'(x_0)p_{3|x_0}\nonumber\\
=\Phi''(x_0)
(x-x_0)+\E_3'(x_0)+\E_3'(x_0)[p_{2|x_0}-p_{1|x_0}].
\label{osh}
\end{gather}
We shall focus on the last term in (\ref{osh}) that determines the
shape of $\Phi'(x)$ as a function of $\beta_{31}$ and $\beta_{32}$.
This term is expanded for $\beta_{31}\approx\beta$,
$\beta_{32}\approx\beta$:
\begin{gather}
b\equiv \E_3'(x_0)[p_{2|x_0}-p_{1|x_0}]\nonumber\\
=
\E_3'(x_0){\sum}_{\alpha={31}, {32}}(\beta_\alpha-\beta)\partial_{\beta_{\alpha}}[
p_{2|x_0}-p_{1|x_0}].
\comment{\Phi''(x_0)={\sum}_{k=2}^3\E_k''(x_0)p_{k|x_0}+ [\E_3'(x_0)]^2\left[
2\frac{\partial g}{\partial\E_2}+\frac{\partial g}{\partial\E_3}
\right] ; g=2p_{2|x_0}+p_{3|x_0}}
\label{bala}
\end{gather}
To work out (\ref{bala}) via (\ref{77}--\ref{7777}) of the main text,
we shall assume for the transition rates $\rho_{i\leftarrow j}$:
\begin{eqnarray}
  \label{eq:222}
  \rho_{ij}(x)=f_{ij}[\,\beta_{ij} (E_j(x)-E_i(x)),\beta_{ij}\,], ~~~ 
\beta_{ij}=\beta_{ji},
\end{eqnarray}
where $f_{ij}[y,\beta]$ holds the detailed balance conditions; see
(\ref{eq:2}) of the main text. Eq.~(\ref{eq:222}) is more general than
its analogue (\ref{eq:22}) in the main text. Combining (\ref{bala})
with (\ref{eq:222}) and with (\ref{77}--\ref{7777}) of the main text,
we get
\begin{gather}
  \label{eq:10}
p_{2|x_0}-p_{1|x_0}=\frac{1}{Z(x_0)} 
[f_{32}(\beta_{32}\E_3(x_0))f_{31}(-\beta_{31}\E_3(x_0))-\nonumber\\
f_{31}(\beta_{31}\E_3(x_0))f_{32}(-\beta_{32}\E_3(x_
0))\,], \\
  b=\frac{\E_3(x_0)\E_3'(x_0)}{Z(x_0)}{\sum}_{\alpha={31,32}}
(\beta-\beta_{\rm \alpha})\psi_{\alpha}[\beta \E_3(x_0)],
\end{gather}
where we denoted $f'_{ij}[y,\beta]\equiv\partial_yf_{ij}[y,\beta]$,
\begin{gather}
\psi_{31}[x]\equiv f'_{31}[x,\beta]f_{32}[-x,\beta]
+f'_{31}[-x,\beta]f_{32}[x,\beta], \\
\psi_{32}[x]\equiv -f'_{32}[-x,\beta]f_{31}[x,\beta]
-f'_{32}[x,\beta]f_{31}[-x,\beta].\nonumber
\end{gather}
Now note the following inequality
\begin{gather}
  \label{eq:14}
f'_{ij}[y,\beta]=\partial_yf_{ij}[y,\beta]\geq 0.
\end{gather}
It means that the transition from a lower energy to a higher energy is
facilitated, if the lower energy increases or the higher energy
decreases. This inequality does follow from the detailed balance [see
(\ref{eq:2}) of the main text], but it still holds for all physical
examples we are aware of.  The inequality implies $\psi_{\rm 31}[\beta
\E_3(x_0)]\geq 0$ and $\psi_{\rm 32}[\beta \E_3(x_0)]\leq 0$. Choosing
$\beta$ in between of $\beta_{31}$ and $\beta_{32}$ we see that
\begin{eqnarray}
  \label{eq:15}
  {\rm sign}[b]={\rm sign}[\E_3(x_0)\E_2'(x_0) (1-\vartheta)]
\end{eqnarray}
Working out the heat-engine condition [see (\ref{kaa}) of the main
text] in the considered order ${\cal O}(|1-\vartheta|)$ and ${\cal
  O}(|x-x_0|)$ we obtain
\begin{eqnarray}
  \label{eq:35}
&& {\E_3(x_0) (1-\vartheta)}\left/{\E_2'(x_0)}\right.> x-x_0>0, \quad {\rm
    or} \\
&& {\E_3(x_0) (1-\vartheta)}\left/{\E_2'(x_0)}\right.< x-x_0<0.
  \label{eq:36}
\end{eqnarray}
Conditions (\ref{eq:15}--\ref{eq:36}) imply that depending on the sign
of $\Phi''(x_0)$ in (\ref{osh}), $\Phi'(x)$ can assume in the vicinity
of $x_0$ only two possible shapes; one of them is shown in Figs. 3 and
4 of the main text. Obviously, neither of them is compatible with
$\Phi''(\x)>-E''_1(\x)$; see (\ref{eq:29}) of the main text.

\subsection{3. Energy dissipation due to external force that generates
  negative friction}

Consider (\ref{eq:111}, \ref{eq:12}) of the main text that we
generalize as follows:
\begin{gather}
  \label{eq:3}
\dot{p}(x,t)=
\frac{1}{\gamma}
\partial_x[p(x,t)\Psi'(x)]+\frac{T}{\gamma}\partial^2_x p(x,t) \\
+\partial_x[p(x,t)G'(x)],
  \label{eq:33}
\end{gather}
where we assume that $x$ couples with a thermal bath at temperature
$T$, 
\begin{eqnarray}
  \label{eq:39}
\gamma>0,  
\end{eqnarray}
is the friction constant, and $G'(x)=\frac{\d}{\d
  x}G(x)$ is an external force. If now we set
\begin{eqnarray}
  \label{eq:31}
  G(x)=-\frac{2}{\gamma}\Psi(x),
\end{eqnarray}
the resulting influence of $G(x)$ and $\Psi(x)$ is equivalent to a
negative friction.

The average energy $\Pi$ dissipated per unit of time due to the external
force $G'(x)$ can be estimated via the change of the free energy
\begin{eqnarray}
  \label{eq:26}
F=\int \d x\, p(x,t)[\Psi(x)  +T\ln p(x,t)],
\end{eqnarray}
of $x$ due to the external part (\ref{eq:33}) 
of the dynamics
\begin{eqnarray}
  \label{eq:27}
  \Pi= \int \d x\, \partial_x[p(x,t)G'(x)]\,[\Psi(x)  +T\ln p(x,t)].
\end{eqnarray}
In the stationary state:
\begin{gather}
  \Pi=- \int \d x\, p(x)G'(x)\,[\Psi'(x) +T\frac{\d }{\d x}
  \ln p(x)]\\
  = \gamma \int \d x\, p(x)[G'(x)]^2,
  \label{eq:30}
\end{gather}
where $p(x)\propto\exp[-(\Psi(x)+\gamma G(x))/T] $. Using
(\ref{eq:31}) we finally obtain:
\begin{eqnarray}
  \label{eq:34}
  \Pi=\frac{1}{\gamma}\int \d x\, p(x)[\Psi'(x)]^2. 
\end{eqnarray}

\end{document}